\documentstyle[11pt, aasms4]{article}
\input tcilatex
\received{December 2000  }
\accepted{to appear in the ApJ July 2001}
\begin{document}

\baselineskip 12 pt

\title{Relativistic Jet Response to Precession and Wave-Wave Interactions}

\author{Philip E. Hardee}  
\affil{Department of Physics \& Astronomy \\ The University of Alabama \\
Tuscaloosa, AL 35487 \\ hardee@athena.astr.ua.edu}

\author {Philip A. Hughes} 
\affil{Astronomy Department \\ University of Michigan \\
Ann Arbor, MI 48109 \\ hughes@astro.lsa.umich.edu} 

\author{Alexander Rosen}
\affil{Armagh Observatory \\ College Hill \\ 
 Armagh, BT61 9DG \\ UK \\ rar@star.arm.ac.uk}

\author {Enrique A. Gomez} 
\affil{Department of Physics \& Astronomy \\ The University of Alabama \\
Tuscaloosa, AL 35487 \\ gomez@hera.astr.ua.edu}

\begin{abstract}

Three dimensional numerical simulations of the response of a Lorentz
factor $2.5$ relativistic jet to precession at three different
frequencies have been performed.  Low, moderate and high precession
frequencies have been chosen relative to the maximally unstable
frequency predicted by a Kelvin-Helmholtz stability analysis.
Transverse motion and velocity decreases as the precession frequency
increases. Although helical displacement of the jet decreases in
amplitude as the precession frequency increases, a helical shock
is generated in the medium external to the jet at all precession
frequencies.  Complex pressure and velocity structure inside the jet is
shown to be produced by a combination of the helical surface and first
body modes predicted by a normal mode analysis of the relativistic
hydrodynamic equations.  The surface and first body mode have different
wave speed and wavelength, are launched in phase by the periodic
precession, and exhibit beat patterns in synthetic emission images.
Wave (pattern) speeds range from $0.41c$ to $0.86c$ but beat patterns
remain stationary.  Thus, we find a mechanism that can produce
differentially moving and stationary features in the jet.

\end{abstract}


\baselineskip 14pt

\section{Introduction}

Relativistic jets, particularly in extragalactic sources, can exhibit
time-dependent curved structures with superluminally moving components,
e.g., 3C~345 (Zensus, Cohen \& Unwin \markcite{zcu}1995) or with both
superluminally moving and much more slowly moving or stationary
components, e.g., M~87 (Biretta, Zhou \& Owen \markcite{bzo}1995;
Biretta, Sparks \& Macchetto \markcite{bsm}1999), 4C~39.25 (Alberdi et
al.\ \markcite{agmm}2000), 3C~120 (Walker et al.\ \markcite{wbu}2001).
Superluminal motions along curved trajectories can be explained by
helical jet models (Hardee \markcite{har87}1987; Steffan et
al.\ \markcite{szk}1995), and helical patterns are the expected result
for Kelvin-Helmholtz and current driven  jet instabilities in
relativistic flows (Birkinshaw \markcite{birk}1991, and references
therein; Appl \markcite{appl}1996; Istomin \& Pariev
\markcite{ip}1996). It has been proposed that a combination of moving
and stationary components can be the result of enhanced features moving
with the jet flow through relatively fixed curved helical structures
with Doppler boosting leading to fixed components where the average
flow comes most nearly towards the line-of-sight (Alberdi et
al.\ \markcite{agmm}2000; Walker et al.\ \markcite{wbu}2001).

Axisymmetric relativistic jet simulations have investigated component
motion by introducing velocity perturbations at the origin and studying
the subsequent evolution.  With repeated velocity perturbations (Gomez
et al.\ \markcite{gmm}1997; Midouszewski, Hughes \& Duncan
\markcite{mhd}1997) the resulting knotty structure moves with the jet
flow.  In a more recent axisymmetric simulation performed by Agudo et
al.\ \markcite{agmi}(2001), a single velocity perturbation generates a
superluminal component associated with the velocity perturbation and
multiple trailing sub-luminally moving accelerating components.  These
simulations address some of the issues involving components moving at
different speeds but do not address the issue of components moving
through stationary features or curved trajectories.  To this end
several 3-D hydrodynamic simulations of relativistic jets have been
performed (Aloy et al.\ \markcite{aim}1999; Aloy et
al.\ \markcite{agi}2000; Hughes et al.\ \markcite{hmd1}1999,
\markcite{hmd2}2001) in which the effect of perturbation induced
asymmetries on jet propagation were investigated.  However, these
studies did not address the issue of internal jet structure associated
with asymmetric fluid motion and the component motions that might be
observed inside the jets.

In this paper we present fully 3-D hydrodynamic simulations along with a
detailed analysis of the time-dependent structures that develop in the
jet for three different jet precession frequencies but with no velocity
variation other than the relatively small transverse velocity induced
by the jet precession.  The numerical techniques are described in \S
2.  Results and analysis of jet structures seen in the simulations are
presented in \S 3 and \S 4.  The potential observable features
resulting from these structures are shown in \S 5, and in \S6 we
summarize and discuss some of the implications of our results.

\vspace{-0.7 cm}

\section{CFD Simulations}

\vspace{-0.3 cm}

\subsection{Solution of the Euler Equations}

We assume an inviscid and compressible gas, and an ideal equation of
state with constant adiabatic index. We use a Godunov-type solver which
is a relativistic generalization of the method of Harten, Lax, \&
Van Leer \markcite{hlv}(1983), and Einfeldt \markcite{ein}(1988), in
which the full solution to the Riemann problem is approximated by two
waves separated by a piecewise constant state.  We evolve the mass density
$R$, the three components of the momentum density $M_x$, $M_y$ and
$M_z$, and the total energy density $E$ relative to the laboratory
frame.

Defining the vector
\begin{equation}
U = (R,M_x,M_y,M_z,E)^{T} ,
\end{equation}
and the three flux vectors
\begin{equation}
F^x = (R v^x,M_x v^x + p, M_y v^y, M_z v^z, (E + p)v^x) ^{T } ,
\end{equation}
\begin{equation}
F^y = (R v^y,M_x v^x, M_y v^y + p, M_z v^z, (E + p)v^y) ^{T } ,
\end{equation}
\begin{equation}
F^z = (R v^z,M_x v^x, M_y v^y, M_z v^z + p,(E + p)v^z) ^{T } ,
\end{equation}
the conservative form of the relativistic Euler equation is
\begin{equation}
{\partial{ U}\over \partial{t}} + {\partial\over \partial{x} } (F^x) +
{\partial\over \partial y} (F^{y})+ {\partial\over \partial z} (F^{z}) = 0 .
\end{equation}
The pressure is given by the ideal gas equation of state $ p = (\Gamma
- 1) (e - n)$ where here and in what immediately follows we have set $c
= 1$. The Godunov-type solvers are well known for their capability as
robust, conservative flow solvers with excellent shock capturing
features.  In this family of solvers one reduces the problem of
updating the components of the vector $U$, averaged over a cell, to the
computation of fluxes at the cell interfaces. In one spatial dimension
the part of the update due to advection of the vector $U$ may be
written as
\begin{equation}
{U^{n+1}}_{i} = {U^{n}}_{i} - \frac{\delta t}{\delta x} (F_{i + \frac{1}{2} } -
F_{i - \frac{1}{2}}) .
\end{equation}
In the scheme originally devised by Godunov \markcite{god}(1959), a
fundamental emphasis is placed on the strategy of decomposing the
problem into many local Riemann problems, one for each pair of values
of $U_{i}$ and $U_{i+1}$ to yield values which allow the computation of
the local interface fluxes $F_{i+\frac{1}{2}}$.  In general, an initial
discontinuity at $i+\frac{1}{2}$ due to $U_{i}$ and $U_{i+1}$ will
evolve into four piecewise constant states separated by three waves.
The left-most and right-most waves may be either shocks or rarefaction
waves, while the middle wave is always a contact discontinuity. The
determination of these four piecewise constant states can, in general,
be achieved only by iteratively solving nonlinear equations.  Thus the
computation of the fluxes necessitates a step which can be
computationally expensive.  For this reason much attention has been
given to approximate, but sufficiently accurate, techniques.  One
notable method is that of Harten, Lax, \& Van Leer \markcite{hlv}(1983;
HLL), in which the middle wave, and the two constant states that it
separates, are replaced by a single piecewise constant state.  One
benefit of this approximation, which smears the contact discontinuity
somewhat, is to eliminate the iterative step, thus significantly
improving efficiency.  However, the HLL method requires accurate
estimates of the wave speeds for the left- and right-moving waves.
Einfeldt \markcite{ein}(1988) analyzed the HLL method and found good
estimates for the wave speeds. The resulting method combining the
original HLL method with Einfeldt's improvements (the HLLE method), has
been taken as a starting point for our simulations.  In our
implementation we use wave speed estimates based on a simple
application of the relativistic addition of velocities formula for the
individual components of the velocities, and the relativistic sound
speed, $a$, assuming that the waves can be decomposed into components
moving perpendicular to the three coordinate directions.

In order to compute the pressure $p$ and sound speed $a$ we need the
rest frame mass density $n$ and energy density $e$.  However, these
quantities are nonlinearly coupled to the components of the velocity as
well as to the laboratory frame variables via the Lorentz
transformation:
\begin{equation}
R = \gamma n ,
\end{equation}
\begin{equation}
M^{x} = \gamma^2 ( e + p ) v^{x} ,
\end{equation}
\begin{equation}
M^{y} = \gamma^2 ( e + p ) v^{y} ,
\end{equation}
\begin{equation}
M^{z} = \gamma^2 ( e + p ) v^{z} ,
\end{equation}
\begin{equation}
E = \gamma^2 ( e + p ) - p ,
\end{equation}
where $\gamma = ( 1 - v^2 )^{-1/2}$ is the Lorentz factor and $v^2 =
(v^{x})^2 + (v^{y})^2 + (v^{z})^2$. When the adiabatic index is
constant it is possible to reduce the computation of $n$, $e$, $v^{x}$,
$v^{y}$ and $v^{z}$ to the solution of the following quartic equation:
\begin{eqnarray}
& & \Bigl\lbrack\Gamma v \left(E-Mv\right)-M  \left(1-v^2\right)\Bigr\rbrack^2
 \nonumber \\
& & -  \left(1-v^2\right)v^2\left(\Gamma-1\right)^2R^2=0 ,
\end{eqnarray}
where $M^2 = (M^{x})^2 + (M^{y})^2 + (M^{z})^2$. This quartic is solved
at each cell several times during the update of a given mesh using
Newton-Raphson iteration.

Our scheme is generally of second order accuracy, which is achieved by
taking the state variables as piecewise linear in each cell, and
computing fluxes at the half-time step. However, in estimating the
laboratory frame values on each cell boundary, it is possible that
through discretization, the lab frame quantities are unphysical -- they
correspond to rest frame values $v>1$ or $p<0$. At each point where a
transformation is needed, we check that certain conditions on $M/E$ and
$R/E$ are satisfied, and if not, recompute cell interface values in the
piecewise constant approximation.  We find that such `fall back to
first order' rarely occurs.

\vspace{-0.7 cm}

\subsection{Adaptive Mesh Refinement}

The relativistic HLLE (RHLLE) method constitutes the basic flow
integration scheme on a single mesh.  We use adaptive mesh refinement
(AMR) in order to gain spatial and temporal resolution.  The AMR
algorithm used is a general purpose mesh refinement scheme which is an
outgrowth of original work by Berger \markcite{ber}(1982) and Berger
\& Colella \markcite{bac}(1989).  The AMR method uses a hierarchical
collection of grids consisting of embedded meshes to discretize the
flow domain.  We have used a scheme which subdivides the domain into
logically rectangular meshes with uniform spacing in the three
coordinate directions, and a fixed refinement ratio of $\times 3$. The
AMR algorithm orchestrates i) the flagging of cells which need further
refinement, assembling collections of such cells into meshes; ii) the
construction of boundary zones so that a given mesh is a self-contained
entity consisting of the interior cells and the needed boundary
information; iii) mechanisms for sweeping over all levels of refinement
and over each mesh in a given level to update the physical variables on
each such mesh; and iv) the transfer of data between various meshes in
the hierarchy, with the eventual completed update of all variables on
all meshes to the same final time level. The adaption process is
dynamic so that the AMR algorithm places further resolution where and
when it is needed, as well as removing resolution when it is no longer
required.  Adaption occurs in time, as well as in space: the time step
on a refined grid is less than that on the coarser grid, by the
refinement factor for the spatial dimension. More time steps are taken
on finer grids, and the advance of the flow solution is synchronized by
interleaving the integrations at different levels. This helps prevent
any interlevel mismatches that could adversely affect the accuracy of
the simulation.

In order for the AMR method to sense where further refinement is
needed, some monitoring function is required.  In general we find that
recognizing the presence of significant mass density gradients, contact
surfaces, or strong shear is effective.  The choice of function is
determined by the part of the flow that has the most significance in a
given study.  For the simulations presented here the first two methods
were employed, with the location of contact surfaces being found by
comparing cell-to-cell pressure differences with cell-to-cell
laboratory frame mass density differences. Since the tracking of shock
waves is of paramount importance, a buffer ensures the flagging of
extra cells at the edge of meshes, ensuring that important flow
structures do not `leak' out of meshes during the update of the
hydrodynamic solution. The combined effect of using the RHLLE single
mesh solver and the AMR algorithm results in a very efficient scheme.
Where the RHLLE method is unable to give adequate resolution on a
single coarse mesh the AMR algorithm places more cells, resulting in an
excellent overall coverage of the computational domain.

\vspace{-0.7 cm}

\subsection{Code Validation and Setup}

The code has been validated using 1-D relativistic shock tube, 3-D
relativistic shock reflection and 3-D relativistic blastwave trials
(Hughes, Miller \& Duncan \markcite{hug}2001).  Furthermore, the
solver employed in the current code is a direct extension to 3-D (with
a recast from Fortran 77 to Fortran 90) of the solver described by
Duncan \& Hughes \markcite{dah}(1994). Evidence for the accuracy and
robustness of that code comes from, in addition to its application to
test problems, the general agreement between studies performed with
that code and with independently constructed codes (e.g., Mart\'\i\ et
al.\ \markcite{ma3}1997).

In the simulations performed here, a `pre-existing' jet flow is
established across the computational volume, to represent the case in
which a leading Mach disk and bow shock have passed, leaving a flow in
pressure balance with a low density external (cocoon) medium -- the
shocked jet material. For all simulations we take the ratio of rest
frame densities to be $\rho_{\rm j}/\rho_{\rm x}=10.0$ where $\rho$ is
the rest frame mass density. The jet flow has $v_{\rm j} = 0.9165c$ and
$\gamma \equiv \left(1-\beta^2 \right)^{-1/2} = 2.5$. The value of the
pressure (and thus the sound speeds inside and outside the jet) is
adjusted to yield a generalized Mach number for the jet of
$\gamma\beta/\gamma_{\rm s}\beta_{\rm s}=8$.  Here $\gamma_{\rm s}
\equiv \left(1-\beta_{\rm s}^2 \right)^{-1/2}$ where $\beta_{\rm s}
\equiv a/c$ and the sound speed, $a$ is given by
\begin{equation}
a \equiv \left[ \frac{\Gamma p}{\rho +\frac \Gamma {\Gamma -1}\frac {p}{c^2}}
\right] ^{1/2}\text{ .} 
\end{equation}
With $\Gamma = 5/3$ as the adiabatic index, the relevant sound speeds
are $a_{\rm x} = 0.6121c$ and $a_{\rm j} = 0.2753c$.

The computational domain is $ 8R_{\rm j} \times 8R_{\rm j} \times 41
R_{\rm j}$, with outflow boundary conditions imposed on all surfaces
except the inflow plane, $z=$~constant.  The inflow plane involves
cells cut by the jet boundary for which state variables must be
established through a volume-weighted average of the internal and
external values. To avoid a `leakage' of jet momentum into the ambient
material, fixed, initial values are used across that entire boundary
plane at every time step.  Schlieren-like images of the pressure, in
which the gradient is rendered on an exponential map, provide a way of
exploring flow structures with a large range of linear scale and
amplitude; an inspection of these images at the last time step of the
computations described here reveals that the computations are free of
any effects due to reflection from the domain boundaries.  Three levels
of refinement were admitted, and as the entire jet is fully refined by
the AMR algorithm initially there are 27 of the finest level cells
across the jet diameter, $2R_{\rm j}$; outside the jet courser cells
are employed initially.

A precessional perturbation is applied at the inflow by imposing a
transverse component of velocity with $v_{\perp}=0.01v_{\rm j}$.
Simulations were performed with precessional perturbations $\omega
R_{\rm j}/v_{\rm j} =$ (A) $0.40$, (B) $0.93$, (C) $2.69$. The
simulations were halted after $\sim 44$ light-crossing-times for the
jet radius, before development of structure had reached a quasi-steady
state across the entire computational grid, when the AMR algorithm
required refinement that exceeded the available computer memory.
 
\vspace{-0.7 cm}

\section{Simulation Results}

Prior to the data analysis we reduced data from all of the refined AMR
``patches" in each relativistic simulation into uniformly spaced
data.   Each simulation contained roughly 300 patches at termination,
with most patches at the finest grid resolution and occurring closer to
the jet than the outer boundary.  Since we wanted to observe the radial
and azimuthal response within the jet at the finest resolution
possible, our uniform grid along the transverse axes is at the finest
resolution of the AMR simulations (13.5 zones/R$_{\rm j}$).  To reduce
the size of the final data sets we have used a moderate scale grid (4.5
zones/R$_{\rm j}$) along the jet axis in our uniformly spaced data.
The lower resolution along the jet axis allows about 10 zones across
2R$_{\rm j}$ features in the axial direction; this length is similar to
the wavelengths seen in the highest frequency simulation.
 
We evaluate simulation (theory) results quantitatively by taking 1-D
slices through data cubes parallel to the $z$-axis at different radial
distances along the transverse $x$-axis ($y$-axis for the theory) as
indicated in Figure 1. For the simulation results shown in Figure 2
the $v_x$ and $v_y$ velocity components correspond to radial, $v_r$,
and azimuthal, $v_{\phi}$, velocity components in cylindrical
geometry.  Axial velocities near to the jet axis show that the
simulations are fully evolved out to about $30R_{\rm j}$ for the low
and moderate frequency simulations and out to about $20R_{\rm j}$ for
the high frequency simulation.  
\begin{figure} [ht!]
\vspace {4.0cm}
\includegraphics{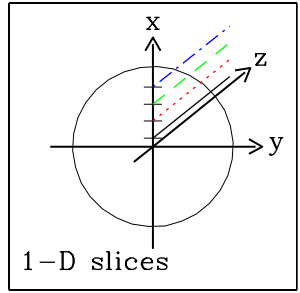}
\caption
{ \baselineskip 12pt
1-D slices through the simulation data cubes parallel to the jet axis
are located at $x/R_{\rm j} =$ (solid line) $0.11$, (dotted line)
$0.33$, (dashed line) $0.55$, (dashed \& dotted line) $0.77$.  Similar
1-D slices through theoretical data cubes are located at similar points
on the $y$-axis.}
\end{figure}

\begin{figure} [h!]
\vspace {10.0cm}
\includegraphics{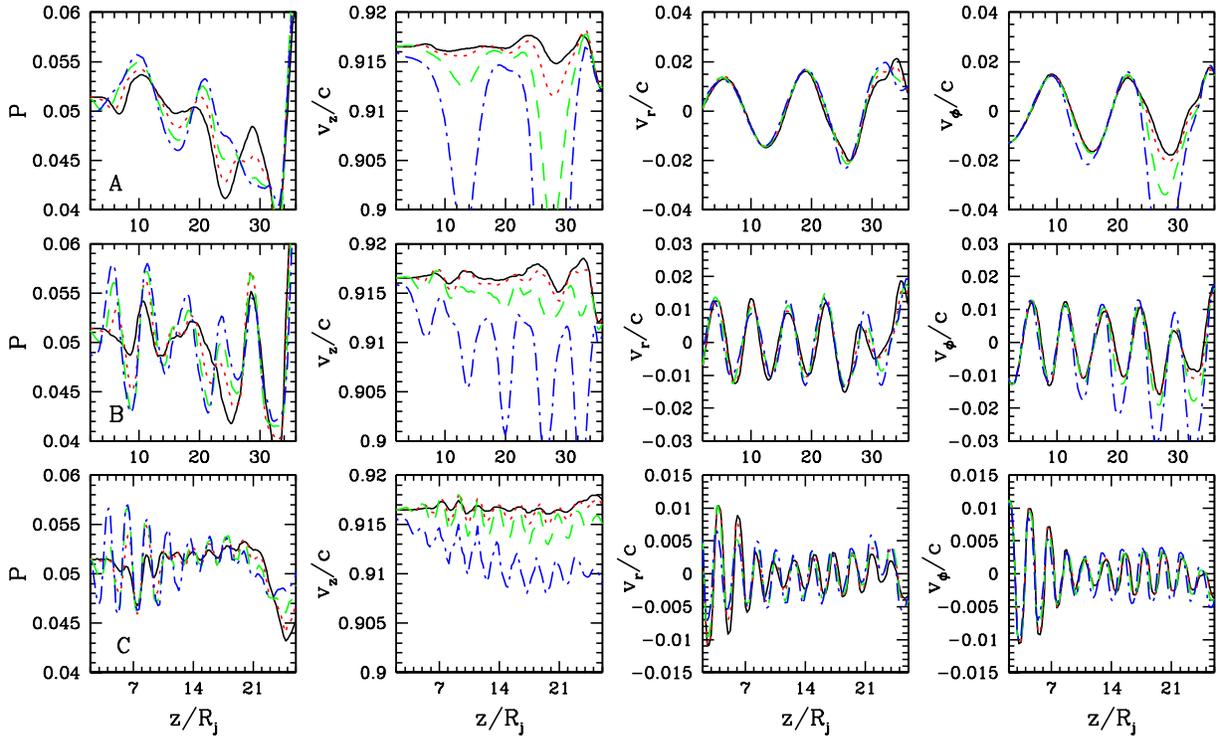}
\caption
{ \baselineskip 12 pt
Pressure, axial velocity ($v_z$), radial velocity ($v_x$), and
azimuthal velocity ($v_y$) along the 1-D slices indicated in
Figure 1 for the three simulations. Line types are as in Figure 1.}
\end{figure}

Slowing of jet material resulting from surface effects is readily
apparent.  Surface effects penetrate more deeply into the jet as the
precession frequency decreases, e.g., large dips in $v_z$ and
$v_{\phi}$ velocity components are observed closer to the jet axis for
lower precession frequency. A dominant wavelength of $\lambda /R_{\rm
j} \sim$ (A) 14, (B) 6, (C) 2 is revealed in the axial velocity plots
in the 1-D slices farthest from the jet axis and in the transverse
velocity plots.  The basic helical nature of the structure induced by
precession is graphically illustrated by the out of phase oscillation
in the transverse velocity components.

In all simulations the pressure structure shows oscillation clearly
related to helical motion in 1-D slices farthest from the jet axis, but
the structure near to the jet axis can be considerably more complicated
-- seen mostly in the low and moderate frequency simulations.  The
maximum pressure fluctuations around the local mean are less than
$\pm15\%$ and are smaller near to the jet axis and in the high
frequency simulation.  Growth in the transverse velocity components,
indicative of unstable helical wave growth, is seen only in the low
frequency simulation.  In the moderate frequency simulation the
transverse velocity component's amplitude remains approximately
constant.  In the high frequency simulation the transverse velocity
components show an initial rapid decline to a plateau.   A subtle
change in the radial velocity structure occurs at $z/R_{\rm j} \sim 10$
in the high frequency simulation.  The change, seen in theoretical fits
(see \S 4), occurs in the amplitude of the 1-D radial velocity slice near
jet center relative to the amplitude of the 1-D slice near the jet
surface.  At larger distance the amplitude near to the center is less
than that at the surface but at smaller distance the amplitude at
center is larger.  A relatively long length scale amplitude modulation
of the plateau in the $v_{\phi}$ velocity component seen beyond
$z/R_{\rm j} \sim 12$ in theoretical fits (see \S 4) may be present to a
small degree in the simulation data.  As we shall see in \S 4,
subtleties like this are indicative of the presence of multiple wave
modes.

Figure 3 shows a spatial Fourier analysis of the 1-D pressure, axial
velocity, and radial velocity slices shown in Figure 2 but windowed
from $1.888\le z/R_{\rm j} \le 30.111$. 
\begin{figure} [h!]
\vspace {9.2cm}
\includegraphics{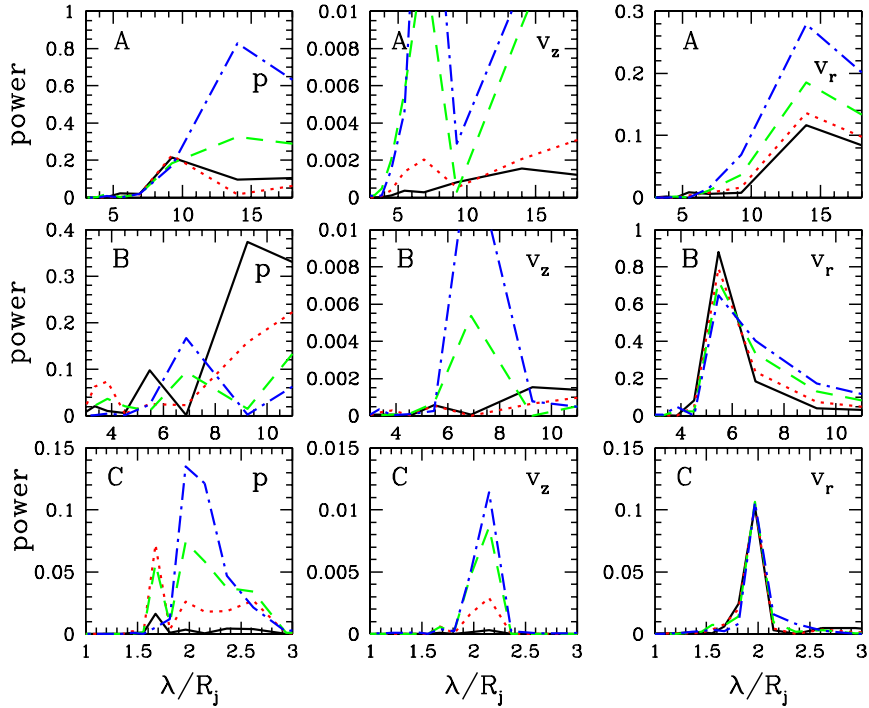}
\caption
{ \baselineskip 12 pt
Spatial FFT of the pressure ($p$), axial velocity ($v_z$), and radial
velocity ($v_r$) along the 1-D slices shown in Figure 2 for the three
simulations.  Line types are identical to Figure 2. Power associated
with velocities has been normalized relative to the maximum radial
velocity power and power associated with pressure has been normalized
relative to the maximum in the pressure.  Off the scale power is not to
scale.}
\end{figure}

All velocity power amplitudes are similarly normalized with the
exception of those off the scale (included so power peaks can be
seen).  Pressure power amplitudes are also similarly normalized.  The
relatively short length of our window results in coarse wavelength
coverage and power is computed at $\lambda/R_{\rm j} \sim$ 28.2, 14.0,
9.3, 6.9, 5.5, 4.5, 3.8, 3.3, 2.9, 2.6, 2.4, 2.1, 2.0, 1.8, 1.7, etc.
We expect power peaks to fall at the wavelength closest to the true
wavelength.  However, various tests indicated that shift in the
location of power peaks to an adjacent wavelength bin could be caused
by amplitude changes in the pressure and velocity oscillations, and
also that the location of power peaks and their amplitude are somewhat
sensitive to the window location.  Still, it is apparent that the power
distribution depends on the radial location within the jet. In general,
power peaks in the radial velocity most nearly indicate the true
wavelength.  To a certain extent these power peaks in the radial
velocity are accompanied by similar power peaks in the pressure and
axial velocity at radial locations $r/R_{\rm j} = 0.55$ \& $0.77$.  At
radial locations $r/R_{\rm j} = 0.11$ \& $0.33$ power peaks in the
pressure can be identified with similar peaks found in the axial
velocity, but there are significant differences.  Study of 1-D cuts
through the data cubes and comparison with theoretical predictions (see
\S 4) lead to the following general conclusions:

\noindent
(1) A small amount of power in the radial velocity at wavelengths below
the power peak suggests the presence of a second transverse
oscillation.   The effect is most evident at radial locations $r/R_{\rm
j} = 0.11$ \& $0.33$. Power at these shorter wavelengths increases from
about $1\%$ to about $10\%$ of the peak power as one goes from low
to high frequency simulations.  Power can also be seen to be enhanced
at shorter wavelengths in pressure and axial velocity.

\noindent 
(2) In all simulations a significant power peak in the pressure and
axial velocity, particularly apparent at radial location $r/R_{\rm j} =
0.11$, occurs at $\lambda/R_{\rm j} \sim 9$.  This feature can be
unambiguously identified with a conical pressure wave at the inlet.
This conical pressure wave appears in the 1-D pressure slices at
$r/R_{\rm j} = 0.11$ in Figure 2 as the pressure dip at axial distance
$z/R_{\rm j} \sim 7 - 8$ followed by a pressure rise at $z/R_{\rm j}
\sim 9 - 10$.

Pressure and transverse velocity vectors in planes transverse to the
$z$-axis at locations of $z/R_{\rm j} = 10$ \& $15$ are shown in Figure
4.  In these cross sections the jet flow is into the page. These
locations were chosen as being far enough outwards to be relatively
unaffected by inlet effects (beyond about $5R_{\rm j}$), not so far
outwards as to be beyond the quasi-steady region [beyond (A \& B)
$30R_{\rm j}$ or (C) $20R_{\rm j}$] or where the jet surface layers are
strongly slowed (beyond $20R_{\rm j}$), and to be far enough separated
to see significant changes in the internal structure of the jet. The
counterclockwise precession of the jet is revealed by spiral shock
waves outside the jet with footpoints on low and high pressure regions
on opposite sides of the jet.  In the medium just outside the jet,
fluid flows from high to low pressure regions around the jet
circumference with less flow at higher precession frequency.  The
highest and lowest pressures and the largest transverse velocities lie
at and outside the jet surface, but in general fluid motions in the
external medium remain less than the sound speed.  The existence of the
spiral shocks suggests the potential for significant energy loss from
the jet surface layers, and the development of a significant axial
velocity shear layer and azimuthal velocity effects appears consistent
with such an energy loss.

In order to see more subtle pressure and transverse velocity structure
inside the jet we have applied an axial velocity mask with value $1$ if
$v_z \ge 0.90c$ and $0$  otherwise to the cross sections.  

\begin{figure} [hp!]
\vspace {19.0cm}
\includegraphics{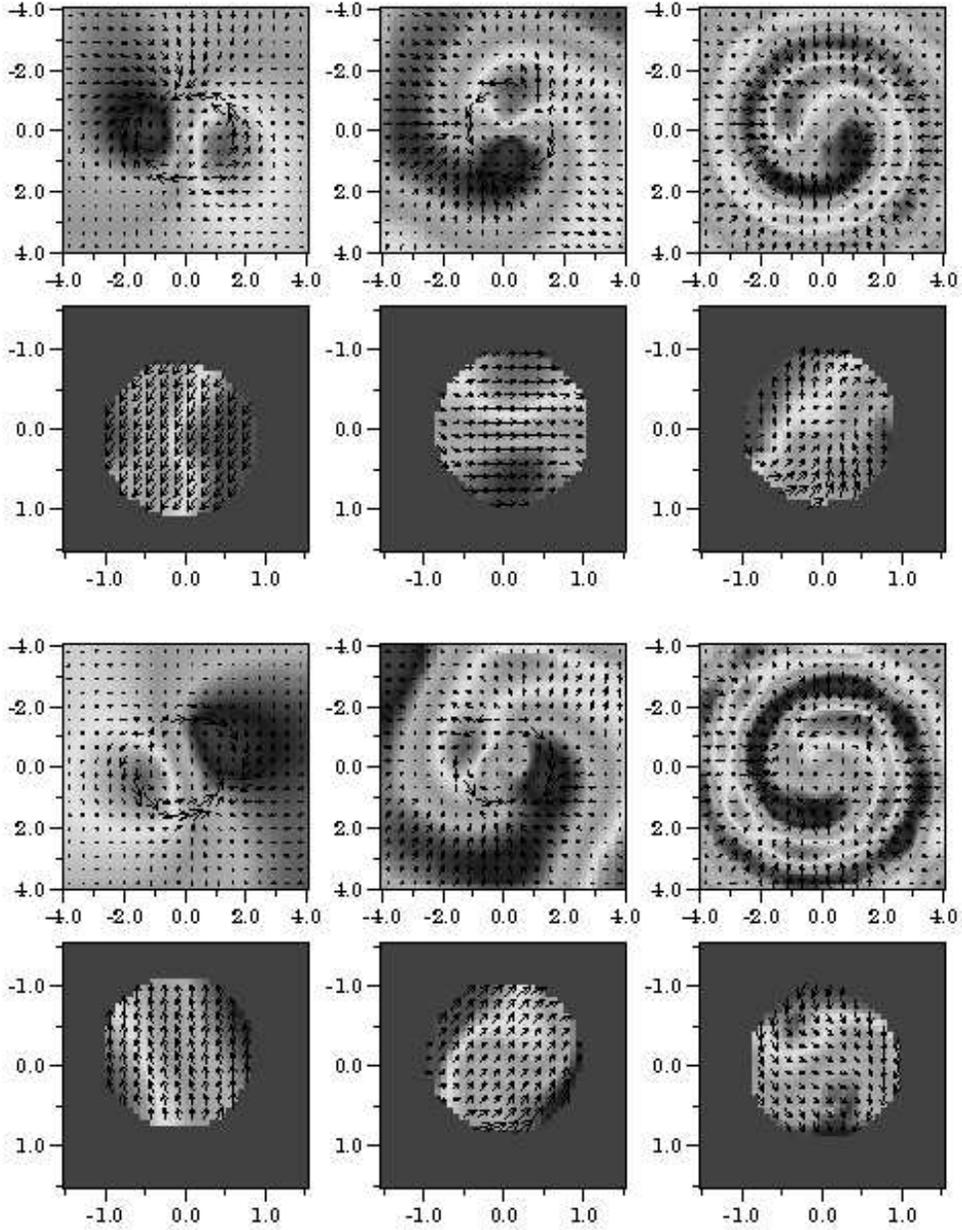}
\caption
{ \baselineskip 12 pt
Transverse cross sections of the pressure structure with transverse
velocity vectors superposed at axial distances $z/R_{\rm j} = $ (top
six panels) $10$ \& (bottom six panels) $15$. Pairs of panels show
transverse structure across the entire computational grid and
transverse structure masked by axial velocity to show only the jet. Low
to high frequency simulation results are shown left to right.  Pressure
color scale and velocity vector length have been adjusted to show
structure and cannot be intercompared quantitatively.}
\end{figure}

The internal pressure structure shows high and low pressure regions
near to the jet surface corresponding to the high and low pressure
regions beyond the jet surface in the unmasked images, but with
additional complicated structure that changes dramatically between the
two locations.  Within the jet, transverse velocities are less than in
the higher sound speed external medium.  In general, transverse
velocity shows jet fluid moving towards the azimuthal location of
maximum jet displacement at a location farther down the jet -- note
that spatial rotation outwards is in the clockwise sense.  This
direction of transverse flow basically proceeds from the leading edge
of the high pressure to leading edge of the low pressure region within
the jet cross sections -- recall that patterns at fixed distance rotate
counterclockwise in a temporal sense.  Flow patterns inside the jet
show some response to internal pressure structure in the high frequency
simulation but are essentially straight across the jet at the lower
precession frequencies.

The complexities in internal jet structure seen in 1-D slices, as
suggested by the power spectrum and by transverse cuts through the jet
indicate that more is going on than can be described by a simple
helical twist of the jet at one wavelength.  In the next section we use
all of the findings above along with a theoretical description of the
normal mode structures of a cylindrical jet to identify the modes
leading to the structures observed in the simulations.

\vspace{-0.7 cm}

\section{Jet Structure}

Suppose we analyze the structures arising in the simulations by
modeling the jet as a cylinder of radius $R_{\rm j}$, having a uniform
density, $\rho _{\rm j}$, and a uniform velocity, $v$. The external
medium (cocoon) is assumed to have a uniform density, $\rho_{x}$, and
to be at rest. The jet is assumed to be in static pressure balance with
the external medium $p_{\rm j}=p_{x}$.  A general approach to
analyzing the time-dependent structures is to linearize the fluid
equations along with an equation of state where the density, velocity,
and pressure are written as $\rho =\rho _0+\rho _1$, $ {\bf v}={\bf
v}_0+{\bf v}_1$ and $p=p_0+p_1$, and subscript 1 refers to a
perturbation to the equilibrium quantity. In cylindrical geometry a
random perturbation of $\rho _1$, ${\bf v}_1$ and $p_1$
can be considered to consist of Fourier components of the form
\begin{equation}
f_n(r,\phi,z)=f_n(r)\exp [i(kz \pm n\phi -\omega t)] 
\end{equation}
where flow is along the $z$-axis, and $r$ is in the radial direction
with the flow bounded by $r=R_{\rm j}$. In cylindrical geometry $k$ is
the longitudinal wavenumber, $n$, is an integer azimuthal wavenumber,
for $n > 0$ the wave fronts propagate at an angle to the flow
direction, the angle of the wavevector relative to the flow direction
is $\theta = tan (n/kR)$, and $+n$ and $-n$ refer to wave propagation
in the clockwise and counterclockwise senses, respectively, when viewed
outward along the flow direction. The dispersion relation describing
the propagation and growth of the ``normal'' modes, $n$, along with the
expressions giving the density, pressure and velocity structure
associated with the normal modes can be found in Hardee et
al.\ \markcite{hhrd}(1998) and Hardee \markcite{har}(2000).

The dispersion relation has been solved for k($\omega $) using root
finding techniques for parameters appropriate to the numerical
simulations.  Numerical solutions to the dispersion relation for the
first two pinch body modes and for the surface and first two helical
body modes appropriate to the simulations are shown in Figure 5.
In general, the body modes are weakly damped in a frequency range just
below where they are not growing (damping rates not shown in the
figure). 
\begin{figure} [ht!]
\vspace {7.0cm}
\includegraphics{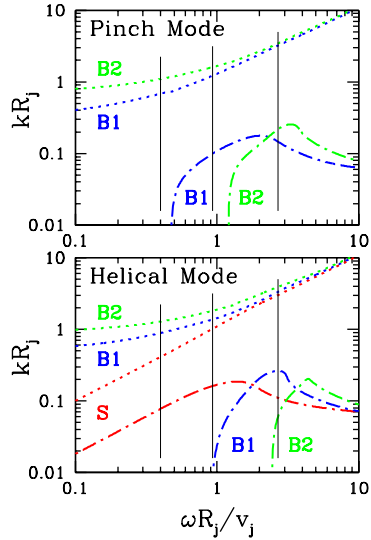}
\caption
{ \baselineskip 12 pt
Numerical solution of the dispersion relation for parameters
appropriate to the simulations for surface (S) wave mode or/and first
two body (B$_1$, B$_2$) wave modes as a function of the angular
frequency. The dotted lines give the real part of the wavenumber,
$k_R$, and the dashed lines give the absolute value of the imaginary
part of the wavenumber, $\left| k_I\right| $.  The vertical lines
indicate the precession frequency in the three numerical simulations.}
\end{figure}

The solutions have comparable maximum growth rates for these
surface and body modes where the spatial growth rate is given by
the imaginary part of the wavenumber, k$_I$.  Wavelengths, wave speeds
and growth (damping) lengths, $\ell \equiv |k_I|^{-1}$, for the helical
modes at the precession frequencies used in the simulations are given
in Table 1.  The wave speed is defined by the real part of the phase
velocity as $v_w = (\omega/k)|_{\rm real}$ and the wavelength is
defined by $\lambda = 2 \pi v_w/\omega$.  We note that non-relativistic
jet simulations have shown wavelength and wave propagation to be given
most accurately by the phase velocity and driving frequency and not,
for example, by using the real part of the complex wavenumber or the
group velocity to determine the wavelength.  The wavelengths given in
Table 1 along with mode structure and amplitudes deduced from the
simulations are used to produce theoretical data cubes.
\vspace{-0.5cm}
\begin{table} [h!]
 \begin{center}
 \caption{Computed Wavelengths, Growth (damping) lengths and Wave speeds  \label{tbl-1}}
 \begin{tabular}{c c c c c c c c c c c} \hline \hline
  {\bf Simulation} & $\omega R_{\rm j}/v_{\rm j}$ 
  & $\lambda_{s}/R_{\rm j}$ & $\ell_s/R_{\rm j}$ & $v_s^w/c$
  & $\lambda_1/R_{\rm j}$ & $\ell_1/R_{\rm j}$  & $v_1^w/c$
  & $\lambda_2/R_{\rm j}$ & $\ell_2/R_{\rm j}$ & $v_2^w/c$ \\  \hline
  A & 0.40 &     14.6  &         12.8        &         0.86     
           &     ~7.1  &        (119)         &         0.41     
           &     ~4.9  &         ---         &         0.29     \\
  B & 0.93 &     ~6.0  &         ~6.2        &         0.82     
           &     ~4.6  &         101         &         0.62     
           &     ~3.5  &         ---         &         0.47    \\
  C & 2.69 &     ~2.1  &         ~9.0        &         0.82     
           &     ~1.9  &          3.8        &         0.74     
           &     ~1.6  &          16.8       &         0.63    \\ \hline
  \end{tabular}
  \end{center}

\end{table}
\vspace {-0.5cm}

Producing theoretical data cubes that yield pressure and velocity
structures similar to pressure and velocity  structures in the
numerical simulations was an iterative process.  We concentrated on
reproducing the general features seen in the simulation 1-D pressure
and velocity slices (Figure 2), and then on the jet's transverse
pressure structure as shown in Figure 4.  These features and structures
along with the simulation power spectra shown in Figure 3 suggested the
need for a pinch mode in addition to helical surface and helical body
modes. This pinch contribution has been modeled by the first pinch body
mode at the wavelength $\lambda_{\rm p}/R_{\rm j} = 8$.  This wavelength is
approximately the length of the conical inlet pressure wave, is equal
to the generalized Mach number, and is consistent with the power
spectra shown in Figure 3. We note that the second pinch body mode
exists but is not growing at this wavelength and higher order pinch
body modes do not exist at this long wavelength.  The 1-D radial
velocity slices, which in the simulations are clearly least affected by
surface effects, proved a good indicator of helical mode amplitudes.
Features in the 1-D pressure slices near to the jet axis provided an
indication of the relative phasing between helical surface and body
modes, and, primarily in the case of the low and moderate frequency
simulations, indicated the presence of pinching triggered by the
conical inlet pressure wave.  Fine tuning of relative phasing between
modes was provided by comparing pressure structure between simulation
and theoretical cross sections.

Results of a combination of helical surface and first body modes along
with (A \& B only) the first pinch body mode are shown as 1-D pressure
and velocity slices in Figure 6.
\begin{figure} [h!]
\vspace {10.0cm}
\includegraphics{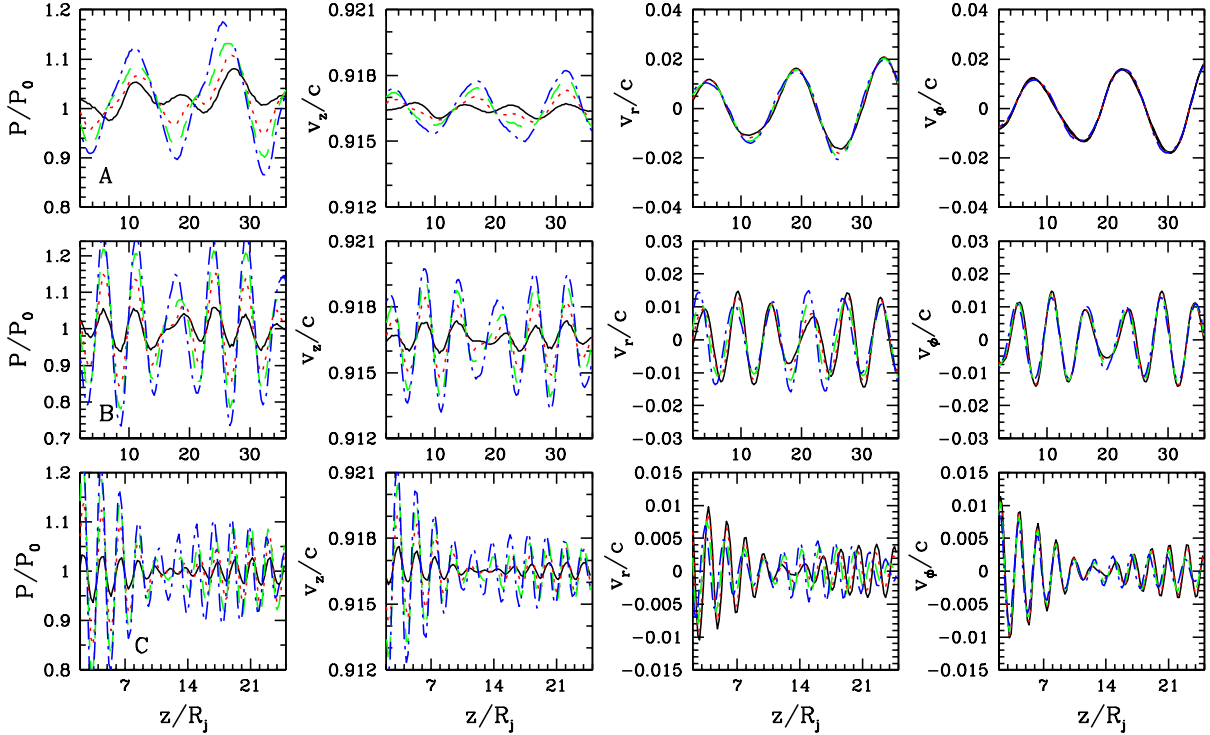}
\caption
{ \baselineskip 12 pt
Pressure, axial velocity ($v_z$), radial velocity ($v_y$), and
azimuthal velocity ($-v_x$) along 1-D slices for theoretical jet models. 1-D
slice locations are now through points along the $y$-axis in Figure 1
but at the same radial locations used for the simulations. Line types
are identical to Figure 2.}
\end{figure}

Qualitatively the pinch mode amplitude grows rapidly to a maximum at
$z/R_{\rm j} = 10$, also the approximate location of the tip of the
conical pressure wave on the jet axis, and declines rapidly at larger
distance, the helical surface mode amplitude grows (A) slowly with
distance, (B) is constant, (C) rapidly declines to a constant value,
and the helical first body mode amplitude (A) declines slowly with
distance, (B) is constant, (C) grows rapidly to a constant value.
Quantitatively amplitudes are input as a maximum jet surface
displacement as a function of $z$, the inlet phase of the helical
surface mode, $\phi_s$, is chosen to match the major radial velocity
oscillations seen in the simulations, and the inlet phase of the
helical body mode, $\phi_1$, is specified relative to the helical
surface mode to produce interference effects and cross section
structure at the appropriate locations. The inlet phase of the pinch
mode is set to produce a maximum on the jet axis at $z/R_{\rm j} =
10$.  The expressions used for the individual modes are:

\noindent
(A)

A$_p=\left\{ 
\begin{array}{cc}
0.00075~z & z<10 \\ 
0.0075\times 2^{-(z/10-1)} & z\geq 10
\end{array}
\right. $

A$_s=0.130+0.005~z$

A$_1=\left\{ 
\begin{array}{c}
0.004 \times e^{-z/120} \\
\phi_1 = \phi_s + 0.26\pi 
\end{array}
\right. $

\noindent
(B)

A$_p=\left\{ 
\begin{array}{cc}
0.0004~z & z<10 \\ 
0.004\times 2^{-(z/10-1)} & z\geq 10
\end{array}
\right. $

A$_s=0.090$

A$_1=\left\{ 
\begin{array}{c}
0.010 \\
\phi _1=\phi _s+0.5\pi 
\end{array}
\right. $

\noindent
(C)

A$_p = 0.0$

A$_s=\left\{ 
\begin{array}{cc}
0.038(1-0.0770~z) & z<10.8 \\ 
0.0065 & z\geq 10.8
\end{array}
\right. $

A$_1=\left\{ 
\begin{array}{cc}
0.00060\times 2^{z/4} & z<13.3 \\ 
0.0060 & z\geq 13.3 \\
\phi _1=\phi _s+1.6\pi &
\end{array}
\right. $

\noindent
where the $A$ and $z$ values above are normalized to the jet radius,
$R_{\rm j}$, and $\phi_{(s,1)}$ are the phase angles at $z = 0$.   At
the higher frequencies jet surface displacement decreases  and the
initial phase difference between helical surface and first body mode
increases.  

In general, the functional forms chosen for the amplitudes are simply a
best fit to what is seen in the simulations.  While amplitude growth
for the pinch mode is chosen to provide a reasonable emulation of the
conical pressure wave at the inlet and has no other physical
significance, subsequent damping may reflect poor coupling between the
conical inlet perturbation and the required structure of this pinch body
mode.  A similar result was found by Hardee et al.\ \markcite{hrhd}(1998) in
axisymmetric relativistic jet simulations. For the first helical body
mode in the low frequency precession simulation we have used the
exponential damping rate predicted by the theory.  Typically the
helical mode wave growth or damping seen in the simulations is not
exponential, e.g., slow linear growth or rapid linear damping of the
surface mode in the low and high precession frequency simulations,
respectively.  This fact implies that amplitudes seen in the
simulations are in the non-linear regime.  Interestingly, the initial
rapid linear damping of the helical surface mode transverse velocity
components in the high frequency simulation is accompanied by rapid
nearly exponential growth of oscillations in the axial velocity
component.  In the simulation, damping of the helical surface mode
ceases when the axial velocity oscillations reach the level appropriate
to the transverse velocity oscillations predicted theoretically to
accompany the helical surface mode at this rapid precession frequency.

While we are able to match the major features in the 1-D slices shown in
Figure 2 on the innermost 1-D slices,  significant
differences between theoretical and simulation 1-D slices appear in the
outer half of the jet where velocity shear appears in the
simulations.  This velocity shear modifies the radial pressure profile,
the axial velocity profile and the azimuthal velocity profile relative
to the theoretical model.  First, the pressure oscillations in the
simulation do not attain the amplitudes seen in the theoretical
profiles in the outer half of the jet.  Second,  the azimuthal velocity
component oscillations can be greatly exaggerated in the simulations in
the outer half of the jet.  This second effect becomes more pronounced
at lower precession frequency and at the accompanying larger sideways
jet displacement.  Third, a larger indication of beating between
helical surface and first body modes appears in the 1-D theoretical
slices.  Nevertheless, results based on the linear theory appear to provide
reasonable estimates of the pressure and velocity fluctuations observed
in the simulations.

Theoretical transverse velocity and pressure structure of individual
helical surface and first body modes at the wavelengths given in Table
1 are shown in the top six panels in Figure 7.  In these panels jet
flow is into the page and spatial rotation of the patterns down the jet
is in a clockwise sense.  At fixed position temporal rotation of the
patterns is in a counterclockwise sense.  For the surface mode at the
fixed spatial location of the slices, transverse flow is approximately
from the leading edge of the high pressure region to the leading edge
of the low pressure region. Flow patterns for the first body mode at
the longer wavelengths are identical and show flow towards (away from)
the regions that will become high (low) pressure regions farther down
the jet.

\begin{figure} [hp!]
\vspace {19.0cm}
\includegraphics{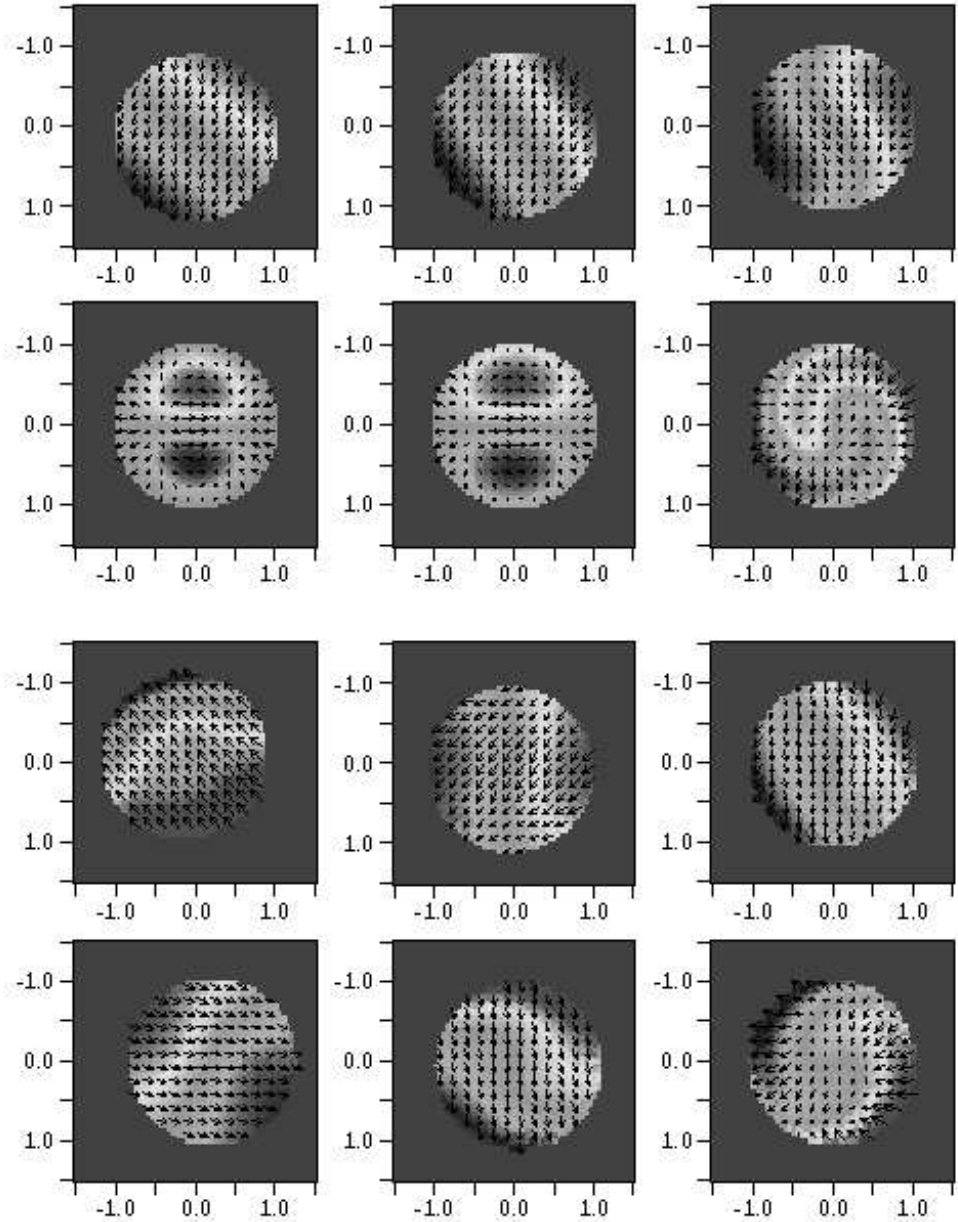}
\caption
{ \baselineskip 12 pt
Transverse cross sections of the pressure structure with transverse
velocity vectors superposed for individual (top row) surface and 1st
body (second row) helical wave modes at (from left to right) low,
moderate, and high precession frequencies.  Transverse cross sections
of the combined modes at axial distances (third row) $z/R_{\rm j} \sim
10$ and (bottom row ) $ \sim 16,15,14$ again at the (from left to
right) low, moderate, and high precession frequencies. Pressure gray
scale and velocity vector length have been adjusted to show structure
and cannot be intercompared quantitatively.}
\end{figure}

Results of the combined modes are shown in the bottom six panels in
Figure 7.  These panels show a theoretical fit to the pressure
structure seen in the cross sections at $z/R_{\rm j} = 10$ \& $15$ in
the simulations.  Because we matched theoretical 1-D slices at radial
positions along the transverse $y$-axis (see Figure 1) to simulation 1-D
slices along the transverse $x$-axis, cross sections through the
theoretical data cubes are rotated by $90^{\circ}$ relative to the
simulation cross sections at comparable positions.  The
theoretical cross sections are taken from a theoretically generated
data cube at axial locations of $z/R_{\rm j} \sim 10$ and $z/R_{\rm j}
\sim $ (A) 16, (B) 15, (C) 14.  Small differences in mode wavelengths
between the theoretical predictions and the simulations lead to the
differences in the outermost location of the best fit.  We are able to
reproduce the basic pressure structure seen in the simulation cross
sections and the results prove that simulation structures are the
result of a combination of normal modes although there are obvious
differences. 

In part, differences between simulation and theory are the result of
the velocity shear layer that occurs in the simulations but is not
included in the theoretical model. We see that transverse flow vectors
are rotated relative to the pressure structure by up to about
$45^{\circ}$ between theoretical cross sections and simulation cross
sections.  Also we find that  pressure structure in the cross sections
is very sensitive to the relative phasing between wave modes.  For
example, the theoretical cross section for simulation A at $z/R_{\rm j}
\sim 10$ requires that the pinch mode provide a central pressure
enhancement while the high pressure region associated with the surface
and first body modes must be very close to $180^{\circ}$ out of phase.
Only this phasing along with appropriate amplitudes provides a pressure
cross section with characteristics similar to that of the simulation.
With the exception of this cross section only the two helical modes
appear important to other cross section structures.  In general the
body mode becomes stronger relative to the surface mode at higher
precession frequency.  Note that transverse velocity structure appears
significantly influenced by the body mode only for the high frequency
case, as in the corresponding simulation cross section.

Figure 8 shows a spatial Fourier analysis of the theoretical 1-D
pressure, axial velocity, and radial velocity slices shown in Figure
6. A window from $z/R_{\rm j} \sim 2 - 30$ was used for low and
moderate precession frequencies and $z/R_{\rm j} \sim 1 - 29$ was used
for the high precession frequency. The length of our theoretical window
and the point spacing is identical to that used for the simulations,
although the normalization is different.   Not surprisingly a Fourier
analysis of the radial velocity slices shows the best correspondence
between theory and simulation as the simulation radial velocity slices
are the least influenced by surface effects.  Note that relative peak
power levels in the radial velocity in the three simulations are
mirrored by the theoretical results. The Fourier analysis of pressure
and axial velocity slices shows much more variability although similar
power peaks are evident in theory and simulation results.  We note that
normalization, particularly for the pressure slices, is very different
for theory and simulation as a result of the large difference in the
amplitude of the pressure fluctuation near to the jet surface.  Other
differences between theory and simulation are the result of different
jet structure both near and far from the inlet in the simulations.  If
factors such as these are considered, the Fourier analysis indicates
good agreement between the theoretical models and the simulations.

\newpage
 
\begin{figure} [ht!]
\vspace {9.2cm}
\includegraphics{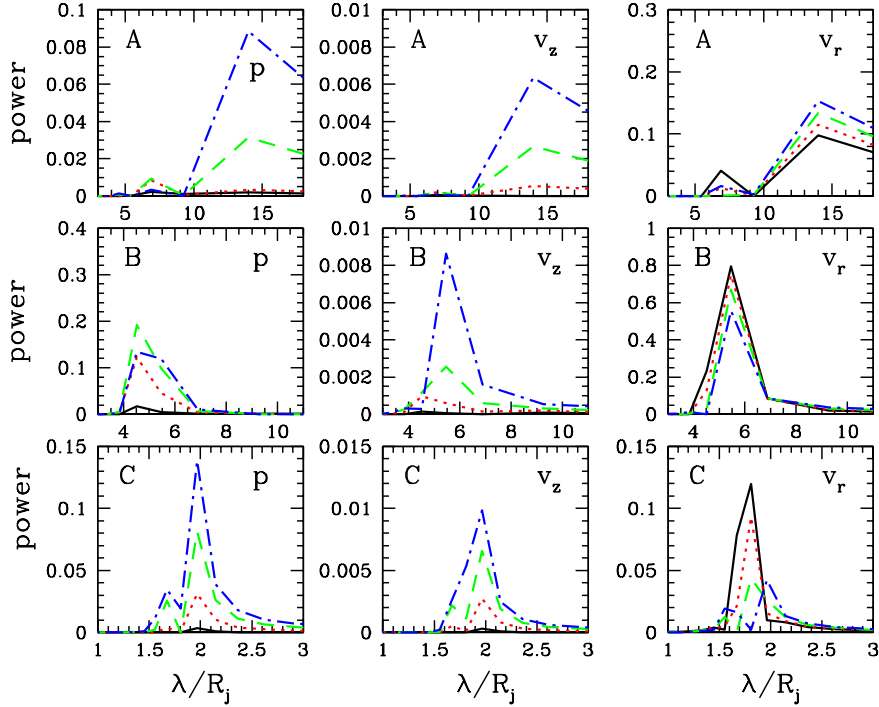}
\caption
{ \baselineskip 12 pt
Spatial FFT of the pressure ($p$), axial velocity ($v_z$), and radial
velocity ($v_r$) along the 1-D slices shown in Figure 6.  Power
associated with velocities and power associated with pressure has been
normalized so that scaling is similar to that shown in Figure 3.}
\end{figure}

\vspace{-0.7cm}

\section{Synthetic Emission Images}

The extent to which complexities in jet structure yield observable
consequences is indicated by the plane of the sky line-of-sight
integrations of p$^2$ shown in Figure 9. Simulation images are
constructed by integrating over only those computational zones in which
$v_z \ge 0.90c$.  Intensities in the different images can be
qualitatively but not quantitatively compared as the scales are not
identical.  Flow and pressure patterns take about $5R_{\rm j}$ to
develop in the simulations so simulation and theoretical images differ
on this scale.  Beyond this distance simulation and theoretical images
are remarkably similar.

The low frequency simulation shows a long wavelength sinusoidal
oscillation at the principal wavelength indicated by the 1-D slices in
Figure 2.  Growth in the amplitude of oscillation is apparent in the
image. Decrease in brightness along the jet in the simulation image is
a manifestation of the average $15\%$ pressure drop between the inlet
and $z/R_{\rm j} = 30$ (see Figure 2).  There is a modest enhancement
in brightness where peaks appeared in the 1-D pressure slices in Figure
2 and at the locations of maximum jet displacement in the plane of the
sky.  The corresponding theoretical image is shifted by the expected
quarter wavelength relative to the simulation image but otherwise
exhibits similar structure.  Modest brightness enhancement at maximum
displacement is a combination of location and shape of the high
pressure region within the jet and line-of-sight effects.  The effect
of pinch overpressure at $z/R_{\rm j} = 10$ is more apparent 

\newpage
\begin{figure} [hp!]
\vspace {18.5cm}
\includegraphics{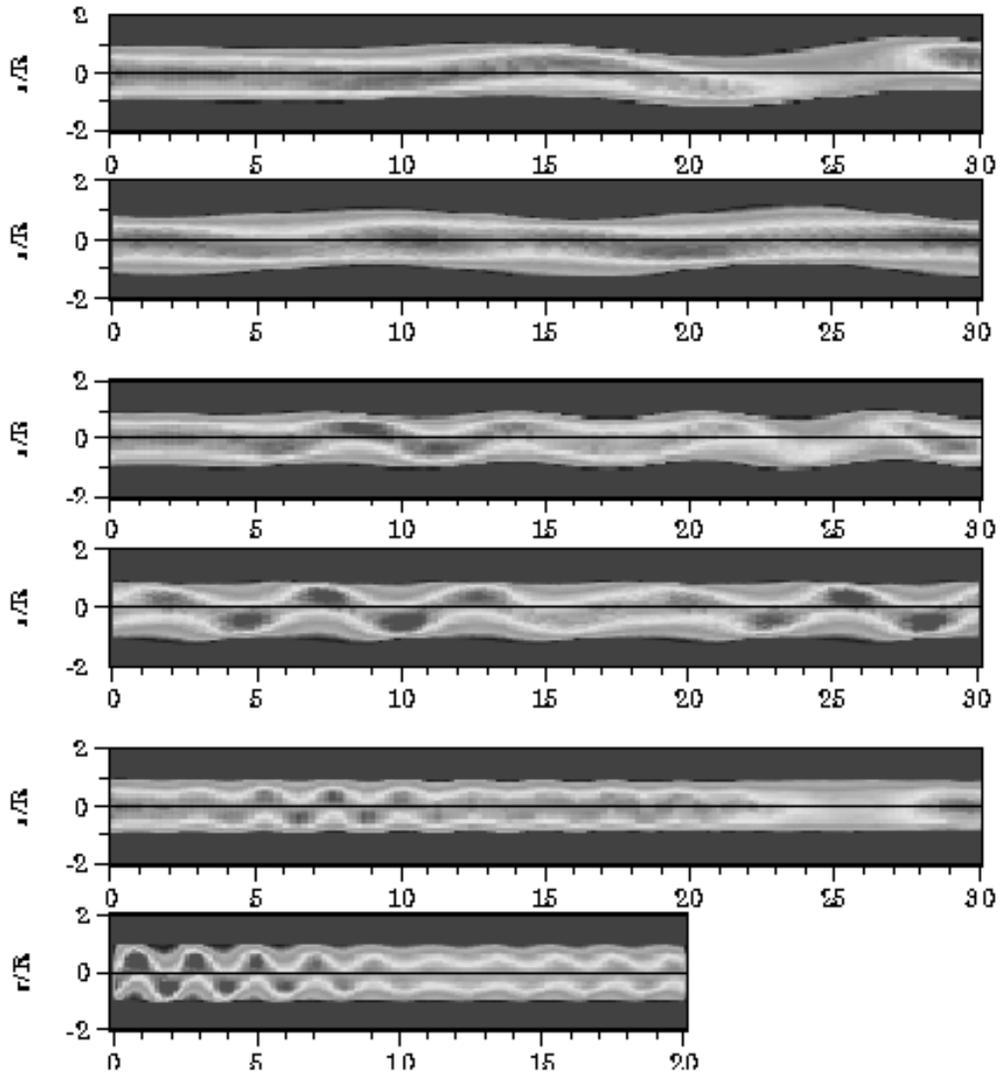}
\caption
{ \baselineskip 12 pt
Line-of-sight integration of $p^2$ through simulation (masked by the
jet velocity to show only the jet) and theoretical data cubes for low
(top two panels), moderate (middle two panels) and high (bottom two
panels) precession frequencies. Color scales have been adjusted to show
structure and cannot be compared quantitatively.}
\end{figure}
\newpage

\noindent
in the theoretical image.  The primary difference between the two
images is a consequence of the jet's axial pressure decline in the
simulation.  The presence of the helical body mode, which is at low
amplitude, has no apparent influence on the appearance of the jet in
simulation or theoretical images.

The moderate frequency simulation shows a sinusoidal oscillation at the
principal wavelength indicated by the 1-D slices, but note the
lengthening in the apparent wavelength only between $z/R_{\rm j} = 14 -
21$.  Some decrease in brightness results from an average $10\%$
pressure drop along the jet in the simulation.  In this simulation
significant brightness enhancement at locations of the maximum
transverse displacement is apparent along much of the jet.  As
suggested by the 1-D velocity slices there is little growth in the
oscillation amplitude.  The corresponding theoretical image shows the
same brightness enhancements at maximum displacement.  There is more
enhancement than in the low frequency images.  It is readily apparent
that the theoretical image reproduces the simulation image structure
between $z/R_{\rm j} = 14 - 21$.  From the theoretical model we deduce
that this structure is the result of interference between helical
surface and first body modes.

The high frequency simulation shows two distinct sinusoidal oscillation
patterns separated by a transition region centered at $z/R_{\rm j} = 12
- 13$ although the oscillation wavelength is nearly identical on either
side of the transition.  A change in transverse velocity structure
occurred just previous to this transition distance and can be seen in
the 1-D slices in Figure 2.   Similar structure is also apparent in the
theoretical image and in the 1-D slices shown in Figure 6.  From the
theoretical model we deduce that this transition region and change in
apparent structure results from decline in the helical surface mode
amplitude and growth in the helical body mode amplitude and
interference between the two modes.  If the theoretical image is
extended to larger distance with constant mode amplitudes a beat
pattern emerges as a result of the slight difference in
wavelength between the two wave modes.  Recall that the beat pattern is
apparent in the 1-D theoretical pressure and velocity slices shown in
Figure 6 and also is suggested in the 1-D simulation velocity slices
shown in Figure 2.

\vspace{-0.7 cm}

\section{Summary and Discussion}

In the simulations, spiral shock waves are driven into the external
medium by jet helicity.  While fluid motions in the external medium
remain less than the sound speed the waves propagate outwards at or
slightly above the sound speed. The footpoints of these waves are tied
to helical ``ripples'' in the jet surface which move axially with
speeds $ 0.86c > v_z^w > 0.74c > a_{\rm x} = 0.61c$.   The azimuthal
motion of the high pressure footpoint at fixed axial position increases
from (A) low to (C) high precession frequency with $v_\phi^w \sim$ (A)
$0.37c$, (B) $0.85c$, and (C) $2.46c$.  In the high frequency
simulation the high pressure footpoint moves around the jet
circumference superluminally and would appear to move across the jet
diameter at about $1.6c$.  The existence of these spiral shocks
suggests the potential for significant energy loss from the jet surface
layers.  In fact the simulations show the development of a significant
axial velocity shear layer and azimuthal velocity effects consistent
with such an energy loss.  Energy loss appears greatest (as judged by
axial and azimuthal velocity effects) at the low precession frequency
where physical displacement of the jet surface is largest and least in
the high precession frequency simulation where physical displacement of
the jet surface is smallest.

Within the jet, fluid motions with respect to the helical pattern speed
are less than the jet sound speed, pressure fluctuates less than $\pm
15\%$ around the local mean and velocity fluctuations in the jet fluid
are much less than the jet sound speed.  Axial velocity fluctuation is
a very small fraction of the relativistic jet speed, and significant
variation in the Lorentz factor does not occur in these relativistic
jet simulations. The small transverse velocity induced by jet helicity
allows for some angular variation in the flow direction. Angular
variation decreases from $\pm 1.25^{\circ}$ in the low frequency
simulation to about $\pm 0.3^{\circ}$ in the high frequency
simulation.  The angular flow variations seen in these simulations are
less than 10\% of the beaming angle given by $1/\gamma$ and given the
relatively small variation in Lorentz factor we would not expect
significant Doppler boosting fluctuations at small angles to the
line-of-sight.  However, we note that larger jet displacements, and
larger pressure and velocity fluctuations would occur at larger
distances outside the computational grid in the low frequency
simulation.  On the other hand, it is likely that the jet
displacements, and pressure and velocity fluctuations in the high
frequency simulation are close to saturation.  Larger jet displacements
and larger pressure and velocity fluctuations cannot be ruled out for
the moderate frequency simulation beyond the length of the
computational grid, although fluctuations remain relatively constant
across the computational grid.

Details of internal jet pressure and velocity structure can be
understood as arising from a combination of the normal modes predicted
by theory.  In general it is possible to provide good estimates of the
velocity and pressure fluctuations in the jet interior but not near the
jet surface where there is significant velocity shear in the
simulations.  Fitting jet structures requires a combination of helical
surface and first body modes with a larger relative amplitude for first
body mode as the precession frequency increases and where first body
mode is more rapidly growing.   Typically the helical mode wave growth
or damping seen in the simulations is not exponential and this fact
implies that amplitudes seen in the simulations are in the non-linear
regime.  Comparison with theory shows that the initial precession
triggers the first helical body mode in addition to triggering the
surface mode even if the body mode is not growing or is weakly damped.
In the simulations some pinching is observed associated with a conical
pressure wave at the inlet. The very oblique pressure wave induced at
the inlet cannot couple strongly to an allowed normal pinch body wave
whose structure includes a much less oblique pressure wave.  Thus, we
would expect damping of the initial perturbation beyond the first
maximum as is suggested by theoretical fits to the simulations. A
similar result was previously found by Hardee et
al.\ \markcite{hrhd}(1998) for axisymmetric jets with much higher
Lorentz factors.
 
In the line-of-sight images the sinusoidal oscillation becomes more
confined to the jet interior as the precession frequency increases, and
the influence of the body mode is enhanced. The high pressure region is
somewhat ribbon like in cross section and this leads to enhancement in
line-of-sight images at the maximum transverse displacement of the high
pressure region. We note that maximum transverse displacement of the
high pressure region is shifted axially relative to the maximum surface
displacement. We observe additional structure within the jet in
addition to the basic sinusoidal oscillation.  The image structure can
only be adequately understood with the modeling capability afforded by
the theory.  In line-of-sight images internal structure arises because
of the configuration and location of the high pressure region
associated primarily with the interacting helical surface and body wave
modes.    The modes are triggered with some initial phase difference
depending on frequency but the phase difference between the modes at
the inlet remains constant.  In the low frequency simulation the body
mode amplitude is small relative to the surface mode amplitude and
there are no observable consequences of wave-wave interaction between
the surface and first body mode.  In the moderate and high frequency
simulations where body mode amplitudes are significant, wave-wave
interaction between surface and body modes does have observable
consequences.  In the moderate frequency simulation, line-of-sight
integrations reveal the effects of constructive and destructive
interference between these wave modes.  In the high frequency
simulation the surface mode declines and body mode grows to comparable
amplitude.  The image reveals interference effects similar to those
seen in the moderate frequency simulation and shows that the body mode
produces effects more confined to the jet interior than the surface
mode.  In any event we see that wave-wave interaction has observable
consequences that are frequency dependent.  Additionally, we find that
the differences between simulation and theoretical models, particularly
in the pressure fluctuations in the outer portion of the jet, do not result
in significant differences in line-of-sight images.

Both simulations and theory suggest potentially interesting wave-wave
interactions resulting from beating between wave modes.  Since the
simulations show a fixed phase difference between modes at the inlet,
the different mode wavelengths result in distinct wave interaction
regions (regions of constructive and destructive interference) that
will remain stationary over time although individual wave patterns will
move through these regions at different wave (pattern) speeds. The
present simulations show that these pattern speeds can be quite
different. Similar wave-wave interactions have been found also in
non-relativistic numerical simulations (Xu, Hardee \& Stone
\markcite{xhs}2000) in a totally different parameter regime relevant to
protostellar jets.  Thus, this effect requires no fine tuning of
parameters and should produce observable consequences on astrophysical
jets in general.   In the present context one might ask if the presence
of moving and fixed components, or quasi-periodic spatial variation in
brightness along jets indicates regions of destructive and constructive
interference between normal modes excited close to the central engine.
In particular, could wave-wave effects lead to the quasi-periodic
spacing of brighter regions in the M~87 jet?  For example, wave-wave
interference could produce particularly complicated structures in a
fixed or slowly moving knot, say in knot D in the M~87 jet, but with
motions within the knot representative of a combination of pattern and
flow speeds with very different apparent velocities.  Future work will
be specifically designed to address these issues.

\vspace {0.2cm}

\noindent
P.\ Hardee, A.\ Rosen and E.\ Gomez acknowledge support from the
National Science Foundation through grant AST-9802955 to the University
of Alabama. P.\ Hughes acknowledges support from the National Science
Foundation through grant AST-9617032 to the University of Michigan.

\end{document}